\documentclass[12pt,preprint]{aastex}

\shorttitle{ROTATION OF HOT HB STARS IN CLUSTERS}
\shortauthors{Recio-Blanco et al\.}

\begin{document}

\def\hst{{\sl HST}}


\title{ROTATION OF HOT HORIZONTAL BRANCH STARS IN THE GLOBULAR CLUSTERS
NGC~1904, NGC~2808, NGC~6093 AND NGC~7078\footnote{Based on
observations with the ESO {\it Very Large Telescope + UVES}, ~at the Paranal Observatory,
Chile}}



\medskip
\author{Alejandra\ Recio-Blanco\footnote{Istituto Nazionale di Astrofisica,
Vicolo dell'Osservatorio 2, I-35122 Padova, Italy; recio@pd.astro.it},
Giampaolo\ Piotto\footnote{Dipartimento di Astronomia, Universit\`a di
Padova, Vicolo dell'Osservatorio 2, I-35122 Padova, Italy;
piotto@pd.astro.it}, Antonio\ Aparicio\footnote{IAC, Via Lactea s/n,
382002 La Laguna Tenerife, Spain; aaj@ll.iac.es}, Alvio\
Renzini\footnote{ESO, Karl-Schwarzschild-Str. 2, D-85748 Garching bei
M$\ddot{u}$nchen, Germany; arenzini@eso.org}}

 

\begin{abstract}
We present high resolution UVES+VLT spectroscopic
observations of 56 stars in the extended horizontal branch (EHB) of the
Galactic
globular clusters NGC~1904, NGC~2808, NGC~6093, and NGC~7078. 
Our data reveal for the first time the presence in NGC~1904 of a sizable
population of fast ($v$sin$i\ge20$ km/s) horizontal branch (HB) rotators,
confined to the cool end of the EHB, similar to that found in
M13.  We also confirm the fast rotators
already observed in NGC~7078.  The cooler stars
(T$_{\rm eff}$ $<$ 11,500 K) in these three clusters show a range of
rotation rates, with a group of stars rotating at $\sim$ 15 km/s or
less, and a fast rotating group at $\sim$ 30 km/s.  Apparently, the
fast rotators are relatively more abundant in NGC~1904 and M13, than
in NGC~7078.  No fast rotators have been identified in NGC~2808 and NGC
6093.  All the stars hotter than T$_{\rm eff}$ $\sim$ 11,500 K
have projected rotational velocities $v$sin$i<$ 12 km/s, but less than
20\% have $v$sin$i<$ 2 km/s.
The connection between photometric gaps in the HB and the
change in the projected rotational velocities is not confirmed
by the new data. However, our data are consistent with a relation between this
discontinuity and the HB jump.
We discuss a number of possibilities for the origin of the 
stellar rotation distribution along the HB. We
conclude that none of them can yet provide a satisfactory explanation
of the observations.
\end{abstract}




\section{Introduction}
\label{intro}

Several unresolved issues in advanced stages of stellar evolution
revolve around the nature of stars in the Horizontal Branch (HB).  In
particular, an increasing number of Globular Clusters (GCs) have been
found to show HB blue tails (Ferraro et al.\ 1998, F98, Piotto et
al.\ 1999, P99), which sometimes extend all the way to the He-burning main
sequence (extended HB stars, EHB), indicating that some
of the stars must have lost (almost) all of their envelope during the
red giant branch (RGB) phase.  We now know that these extremely hot HB stars
can be found in clusters of any metallicity, including in metal rich GCs 
(Rich et al.\ 1997). Yet, the origin of EHBs is still a puzzle.
Stellar evolution models indicate that   
EHB stars are He-core, H-shell burning stars, which have
lost almost their whole envelope during the RGB ascent (Greggio \& Renzini
1990; D'Cruz et al.\ 1996), with a residual envelope mass
$\leq0.02M_\odot$. The problem is that we do not know {\it why}
EHB stars have lost so much mass. Understanding the origin of EHB stars
in GCs has a more general relevance in astrophysics, as these 
very hot stars are now considered the prime contributors to the ultraviolet
emision in elliptical galaxies (Greggio \& Renzini 1990; Brown et al. 2000).

One more puzzling peculiarity is shared
by all the EHBs discovered so far:\ one or more gaps are found in the
stellar distribution along the EHBs, with sections of the HB that are clearly 
underpopulated (Sosin et al.\ 1997, F98, P99), as if mass loss prior to the HB
phase was somewhat {\it quantized}.
Moreover, there is evidence that the GC density favours the appearance of EHBs
(Fusi Pecci et al.\ 1993), hinting that stellar inteactions may favour in some way
the extreme mass loss that is required (Sosin et al. 1997).  Other candidate
scenarios include mixing during either the RGB phase or as a result of a core
helium flash in hot stars (Sweigart 1997; Brown et al. 2001), and
stellar rotation (Peterson et al.\ 1995, P95). In the this paper we will 
focus on stellar rotation rate, investigating whether it is somehow related
to the EHB properties.

Recently, Behr et al.\ (2000a, B00a) have suggested the existence of a discontinuity in
stellar rotation velocity across one of the gaps (at
T$_{\rm eff}$ $\simeq$ 11.000 K) in the EHB of M13. Bluewards of the gap,
all the stars show modest rotations ($v$sin$i < 10$ km s$^{-1}$), while
to the red side of the gap several rapidly rotating stars are found 
(with $v$sin$i$ up to  40 km s$^{-1}$, see also P95).
A similar discontinuity was also found for M15 (Behr et al. 2000b, B00b).
On the other hand, the quick rotation of the M13 HB stars is even more suggestive when
compared with the slower $v$sin$i<20$ km s$^{-1}$ found in clusters without 
an EHB, such as M3 and NGC~288 (P95), suggesting a possible connection 
between rotation and EHBs.

More observations of GCs with different HB morphologies are
needed to understand the role played by rotation on EHB stars. For
this reason, we started an observing campaign with the UVES high resolution
spectrograph at VLT.  In this
paper, we present the results on the rotation of 56 HB stars 
in 4 GCs: NGC~1904 (M79), NGC~2808, NGC~6093 (M80),
and NGC~7078 (M15). The new observation double the number of EHB stars
for which projected rotational velocities have been measured. The 
observations of much larger samples will soon become possible with the 
forthcoming multifiber facilities, such as FLAMES at the VLT.

\section{Observations}
\label{observation}

Our spectra were collected using the VLT-UVES spectrograph 
on Jul 30 - Aug 2 2000, and Jan 19-23 2001. A 1.0 '' slit
width yielded a nominal resolution of R~=~40,000. The UVES blue arm,
with a spectral coverage in the 373 - 499 nm range (where many tens of
metallic lines were expected, and, indeed, identified), has been
used. S/N ratios were always $\geq10$ per resolution element.

As the measurement of $v$sin$i$ by line broadening is inherently
statistical, we have observed more than 10 stars per cluster. Twenty
stars in M79, 11 stars in M15 and 6 stars in M80 were
selected from the HST-WFPC2 images and photometry of the HST-snapshot
program by Piotto et al.\ (2002, cf. http://menhir.pd.astro.it).  
Five additional
stars of M80 were selected from the Johnson {\it VI} photometry by
Rosenberg et al.\ (2000).
Finally, the 11 NGC~2808 stars  were from the CMD
in Bedin et al.\ (2000).  The target stars have
a temperature in the range $8,000\le T_{\rm eff}\le 28,000$K, and are
equally distributed on the two sides of the gap at $T_{\rm
eff}\approx11,000$K present in the EHB of the 4 GCs.  
Figures 1 and 2 (right panels) show the position of the target stars
(full and open circles) on the HB blue tails. 
In general, all the program stars lie in the GC's low crowding
outskirts.  The CCD imaging confirmed the absence of faint companions.
During each observing night, we have also collected high
S/N spectra of a set of field rotational velocity
standards (Peterson et al.\ 1983) , with spectral
types close to that of our program blue HB stars:
HD74721 (3$\pm$3 km/s), HD130095 (3$\pm$3 km/s), HD117880 (12$\pm$3
km/s), HD19445 (13$\pm$3 km/s), and HD109995 (27$\pm$3 km/s).

\section{Data Analysis and Results}
\label{results}

We used the standard IRAF procedures to reduce the spectra.  For the
determination of the projected rotational velocities we used the
cross-correlation technique described by Tonry and Davis (1979).  This
method is well suitable to measure rotational broadening in low
signal-to-noise spectra (Dubath et al.\ 1990).  The analysis procedure
computes (in the Fourier domain) the correlation function of the
object spectrum versus that of a template, fits a gaussian to the
highest peak, and finds the radial velocity and the line broadening
from the peak's central position ($\delta$) and width ($\mu$).
If the Gaussian instrumental width is $\tau$$^{2}$, then a template to
template correlation gives a width 2$\tau$$^{2}$ = $\sigma$$_{o}$,
while a template to object correlation gives a width $\mu$$^{2}$=
$\sigma$$_{o}$$^{2}$+ $\sigma$$^{2}$, where $\sigma$ takes account for
the rotational broadening difference between the template and the
standard.
Finally, the projected rotational velocity $v$sin$i$ can be expressed
by (Melo et al. 2001): $v$sin$i$ $ = A \sqrt{\mu^{2} - 2\tau^{2}} = A
\sigma$,
~ where A is a constant coupling the differential broadening of the
cross-correlation peak, $\sigma$, to the $v$sin$i$ of the stars. The
constant A in the previous equation was found for each rotating
standard star by fitting the relation ($v$sin$i$)$^{2}$ versus
$\mu^{2}$ by a straight line for which the square root of the slope
gives A. The mean value $<$ A $>$ = 1.8 $\pm$ 0.7 km/s was adopted.
Each target spectrum was divided into regions (avoiding hydrogen
lines) of about 40 angstroms, and cross-correlated with the templates,
using the task {\it fxcor} within IRAF. A value of $v$sin$i$ was
calculated for each cross-correlation function, rejecting those
correlations for which the peak's central position does not agree with
the correct stellar radial velocity relative to the template.  In this
way, we avoid errors due to line mismatch or to the lack of spectral
features, specially for the hottest stars and the reddest sections of
the spectra.  Finally, for each program star, we calculated the
weighted mean (typically, from 15 orders) of the $v$sin$i$ values and
the corresponding root mean squares. We used as weight the height of
the cross-correlation peak.  Different templates give consistent 
$v$sin$i$. Possible systematic errors caused by the
dependence of $\sigma$$_{o}$ on other broadening mechanisms, as
microturbulence, are negligible compared to the calculated statistical
error ($\sim$ 3-4 km/s).  More details on the data reduction, and a
table with positions and velocities of the single stars will be
published in a forthcoming paper.

The projected rotation results for the 56 stars in the 4 GCs observed
in this study, and for M13 (from B00a and P95), are shown in Figs.\ 1
and 2. For M15, we plot both our and B00b data. The $v$sin$i$ for two
M15 stars we have in common with B00b are in agreement within the
errors (5 $\pm$ 2 vs. 5.07 $\pm$ 0.24 and 13 $\pm$ 3 vs. 14.88 $\pm$
0.69), showing a consistency between the two sets of observations, and
between the two independent methods adopted for the rotational
velocity measurement.

An estimate of the $T_{\rm eff}$ for each star has been obtained by
comparing the Cassisi et al.\ (1999) models with observed CMDs.  In
particular, we used the {\it uy} Str$\ddot{\rm o}$mgren photometry for
M79 (Grundahl et al.\ 1999, G99); the Johnson {\it UB} for NGC~2808
(Bedin et al.\ (2000), and the HST F439W and F555W for M80 and M15
(Piotto et al.\ 2002).  Hence, there might be some offset in the
temperature scale from GC to GC, in view of the different photometric
systems (the temperatures for the M13, M80 and M15 stars being the
most uncertain ones), but for the purpose of this paper only the
relative position with respect to the HB gaps is relevant.  The gaps
on each GC HB are marked with arrows at the positions suggested by
Hill et al.\ (1996) for M79 (T$_{\rm eff}$ = 9,900K), and F98 and P99
for the other GCs, i.e T$_{\rm eff}$= 11,000K for M80; T$_{\rm eff}$ =
9,000K for M15; T$_{\rm eff}$ = 11,000K for M13. For NGC~2808 we
adopted T$_{\rm eff}$ = 15,900K for the gap position, following Bedin
et al.\ (2000). A second vertical arrow marks the position of the
luminosity ``jump'', taken from G99.  The ``jump'' is a discontinuity
in the Str$\ddot{\rm o}$mgren ({\it u, u$-$y}) locus for which stars
in the range 11,500 $\le T_{\rm eff}\le$ 20,000 K deviate
systematically from (in the sense of appearing brighter and/or hotter
than) canonical zero-age HB models (cf. Fig.\ 2).  The jump seems to
be an ubiquous feature, intrinsic to all HB stars hotter than 11,500 K
(G99).  The jump is also visible in the {\it UB} photometry of Fig.\
1.
    
A first result from this investigation is that in all GCs
{\it all the stars hotter than T$_{\rm eff}$ $\sim$ 11,500 K have
$v$sin$i\le$ 12 km/s}.
This result is based on 116 stars in 5 GCs (including the 31 stars in M13 
and M15 from B00a and B00b, and the 29 stars in M13 from P95).
The measurements obviously provide $v$sin$i$, but for 
an isotropic distribution in rotational axis
large sin$i$ are more likely than small ones, being for
example the probability that sin$i \le 0.25$ about 3\%. Therefore, the bulk
of these stars must be intrinsically slow rotators.
Still, it must be stressed that very few stars (less than 20\%) have
projected rotational velocities below 2km/s (the rotational
velocity of the Sun). Even the ``slow rotators'' have, on
average, $v$sin$i\sim7$ km/s. Only in NGC
2808, 50\% of the stars have projected rotational velocities compatible with
a zero value. The small number of stars do not allow us to conclude
whether the HB stars of this GC are really peculiar or whether
this is just a statistical fluctuation.

In addition, our data reveal for the first time the presence in NGC
1904 of a sizable population of fast HB rotators ($v$sin$i\ge20$ km/s) 
confined to the cool end of the blue HB.  We also confirm
the fast rotators already observed by B00b in M15 (M15).  Among
the cooler stars (T$_{\rm eff}$ $<$ 11,500 K) in these three GCs
there is a range of rotation rates, with a group of stars rotating at
$\sim$ 15 km/s or less, and a fast rotating group at $\sim$ 30 km/s.
Apparently, the fast rotators are relatively more abundant in M79
and M13, than in M15, where only 3 stars out of 22 rotate faster than
15 km/s. In M79 and M13, at least half of our stars cooler than
11,500K are fast rotators. This implies {\it a different intrinsic
distribution in the rotation rates of the EHB stars in these 3 GCs}.  Neither
the $v$sin$i$ distribution in M13 and M79 nor that in M15
are as we could expect from a constant rotation rate and a random
orientation of the axis.

Neither NGC~2808 nor M80 show any fast rotator in our sample.  This
may well result from too few HB stars cooler than 11,500K being included
in our sample. However, there are quite many stars that are cooler than the 
gap, hence this result does not confirm the B00a suggestion that the abrupt 
change in the
rotational velocity distribution was coincident with 
the presence of the EHB gap. All GCs in the present study have an EHB, and 
in all of them there is a HB gap. We were careful in having in all cases
approximately half of the stars on either side of the gap, but Fig.\ 2 
clearly indicates that the presence
of fast rotators {\it is not} related to the presence of the gap.  Indeed,
both in NGC~2808 and M80 there are 8 stars on the cool side of
the gap, and none of them has a rotational velocity exeeding 10 km/s.
The likelihood that the small 
$v$sin$i$ values on the cool side of the gap in NGC~2808 and M80 are due 
to casual almost polar orientation is very small, and
therefore these stars are most likely intrinsic slow rotators.  
Notice also that one star in M15  with $v$sin$i$ = 23$\pm$ 3 km/s is located
significantly bluewards of the gap.

We conclude that {\it the EHB gaps are not related to the abrupt change in
the rotational velocity distribution of the HB stars.}
On the other hand, Fig.\ 2 indicates that all the fast rotators in M15 are 
cooler than $T_{\rm eff}\sim$11,500K, i.e. cooler than the location of the 
G99 jump, rather suggesting a link between the presence of the jump 
and the absence of fast rotators among stars hotter than this temperature.

\section{Discussion.}
\label{discussion}

Along with similar previous studies, the present investigation demonstrates
that HB stars in GCs rotate, and do so much faster than the Sun, in spite of
braking mechanisms having been at work for $\sim $ one Hubble time.
This suggests that either the stars are able to preserve part of their 
angular momentum all the way through very advanced evolutionary stages, or 
that they may re-acquire angular momentum, for example, as a result of tidal interactions 
in the high density environment offered by GCs. 

It is quite possible that the envelope of stars is completely deprived of its
angular
momentum during the RGB phase, when at least half of this envelope is lost 
in a wind. Suffice indeed even a small magnetic field for efficiently transfering
angular momentum from the whole convective envelope - which tends to rotate
solid body - to the wind. However,
the very small, degenerate core may still retain angular momentum,
and on this hypothesis Mengel \& Gross (1976) constructed evolutionary models
with core rotation. In these models rotation has the effect of delaying the 
helium flash until the star reaches a slightly higher luminosity on the RGB.
However small the effect, it would be sufficient to allow more mass to be 
lost by more rapidly rotating stars, offering an explanation for the
origin of the mass dispersion along HB stars (Renzini 1977).
This scenario is in apparent conflict with the observations of stellar 
rotation among EHB stars reported above, because one would expect the fast 
rotators to lose more mass than slow rotators, hence landing at higher 
temperatures on the HB. The opposite is instead observed: the hottest EHB 
stars are all slow rotators, while the fast rotators are found only below
$\sim 11,500$K.

Another embarrassment comes from the mere fast rotation itself: what
rotates is the stellar envelope, which should have lost all its
angular momentum prior to the star beginning its HB phase.  Sills \&
Pinsonneault (2000) proposed models for the angular momentum evolution
in which no magnetic braking takes place, and stars retain some
envelope rotation and a rapidly rotating core during the RGB
phase. Once on the HB, angular momentum redistribution from the core
to the envelope would spin up the envelope.  However, in spite of
being based on assumptions that favor angular momentum retention,
these models fail to predict the high rotation observed below 11,500K,
coupled with the low rotation observed at higher temperatures.  So the
puzzle remains.

Sills \& Pinsonneault (2000) argue that diffusion of heavy elements in the 
hottest stars may prevent angular momentum transfer from the core, due to
the build up of a gradient in mean molecular weight ($\mu$). However, core 
and envelope have quite different $\mu$ anyway, hence diffusion does not
look a viable alternative. Maybe core-envelope angular momentum transfer 
takes place after all, but in stars hotter than $\sim 11,500$K angular
momentum is continuously removed via the radiatively accelerated wind
typical of hot stars. Indeed, the mass loss rate may increase by a large 
factor between $\sim 10,000$ and $20,000$K ( Vink et al.\ 2000),
and may become very small below $\sim 10,000$K, i.e., at the temperatures 
of most fast HB rotators.
While also this remains a speculative solution to the puzzle of the hot slow 
rotators, still the cooler fast rotators seem to require a quite contrived
angular momentum history. Angular momentum extraction from the core should 
be quite inefficient during the RGB (in order to maintain an angular momentum
recevoir), and quite efficient during the HB (in order to promptly spin up
the envelope). We also note that there is no apparent correlation between 
rotational velocity and luminosity {\it distance} from the ZAHB (cf.\ Fig.\ 1 
and 2, right panels), as would result if the timescale of angular momentum 
diffusion were comparable to the HB lifetime. 

As already mentioned, besides all these complications the mere existence of 
fast rotators is not so obvious, given the ample opportunities to lose 
angular momentum along with mass during the RGB phase. 
Soker (1998) has prosposed that fast HB
rotators could have spun up by swallowing close planetary
companions during the RGB phase.  However, no planetary companions have been
found in a very intense search for them in the GC 47 Tuc (Gilliland 
et al. 2000), which seems to exclude this possibility. 
Quite more attractive is the hypothesis of an envelope spin up
as a result of close tidal encounters of RGB
stars with main sequence dwarfs. This scenario is circumstantially 
supported by the noted correlation of the presence of an extended EHB with
the cluster density (Fusi Pecci et al. 1993), with very extended EHBs being 
found almost exclusively in the densest GCs. So, tidal encounters may
account for both enhanced mass loss (e.g. stripping) responsible for the
blue extension of the HB, and for the enhanced rotational velocities of
some HB stars. But for the hot slow rotators another effect must be invoqued.

Mass and angular momentum loss during the EHB phase itself has been 
mentioned above, and the noted coincidence of the drop in rotation with the
luminosity jump may provide support to this hypothesis.
B00a has noted the coincidence between the discontinuity in the rotational 
velocity and the appearance of composition anomalies, most likely due to
diffusion (see also Glaspey et al.\ (1989)). Following Greenstein et al.\ (1967), 
gravitational settling of helium and radiative levitation of metals
can  occur in the stable, non-convective atmospheres
of the hot, high gravity HB stars. This
possibility has been observationally confirmed by Behr et al.\ (1999)
and B00b for M13 and M15, and the jump in the HB luminosity may be
caused by the onset of metal levitation (see also Moehler et al.\ 2000). The fact
that the change in the velocity distribution can be associated to the
jump (instead of to the gap) makes the entire scenario observationally
consistent, especially when also noting that the enhanced surface abundance 
of metals will further boost mass (and angular momentum) loss via
radiation pressure on such elements.
\acknowledgments 
We thank L. R. Bedin, F. Grundahl and A. Rosenberg for providing their
CMDs.  ARB recognizes the support of the {\it INAF}.  GP recognizes
partial support from the {\it MIUR} and from the {\it ASI}.


\clearpage
\begin{figure}
\epsscale{0.9}
\plotone{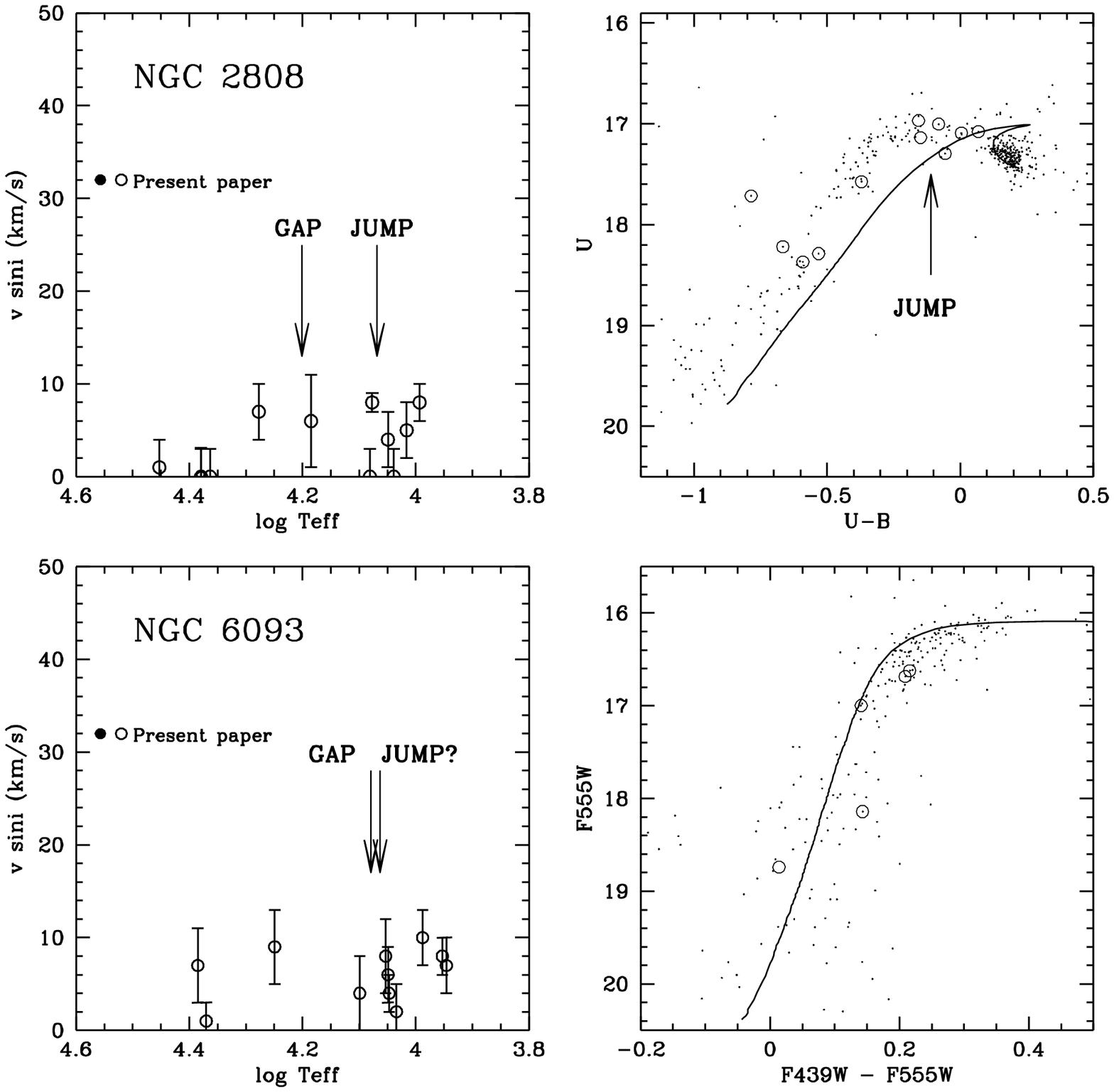}
\figcaption{Left panel: Projected rotational
velocities as a function of the temperature for our targets in NGC
2808 and M80 (open circles).  The vertical arrows indicate the
positions of the ``gaps'' (from F98 and P99) and of the ``jump'' (from G99). 
The position for the jump in M80 has not been
confirmed by observations, yet. We just put the arrow in
correspondence of $T_{\rm eff}=11,500$K where all the HBs of G99
GCs show this feature. Right panel: location of the target stars
in the CMD. The full line represent the best fitting ZAHB from  Cassisi et al.\
(1999).}
\label{slow}
\end{figure}

\begin{figure}
\epsscale{0.9}
\plotone{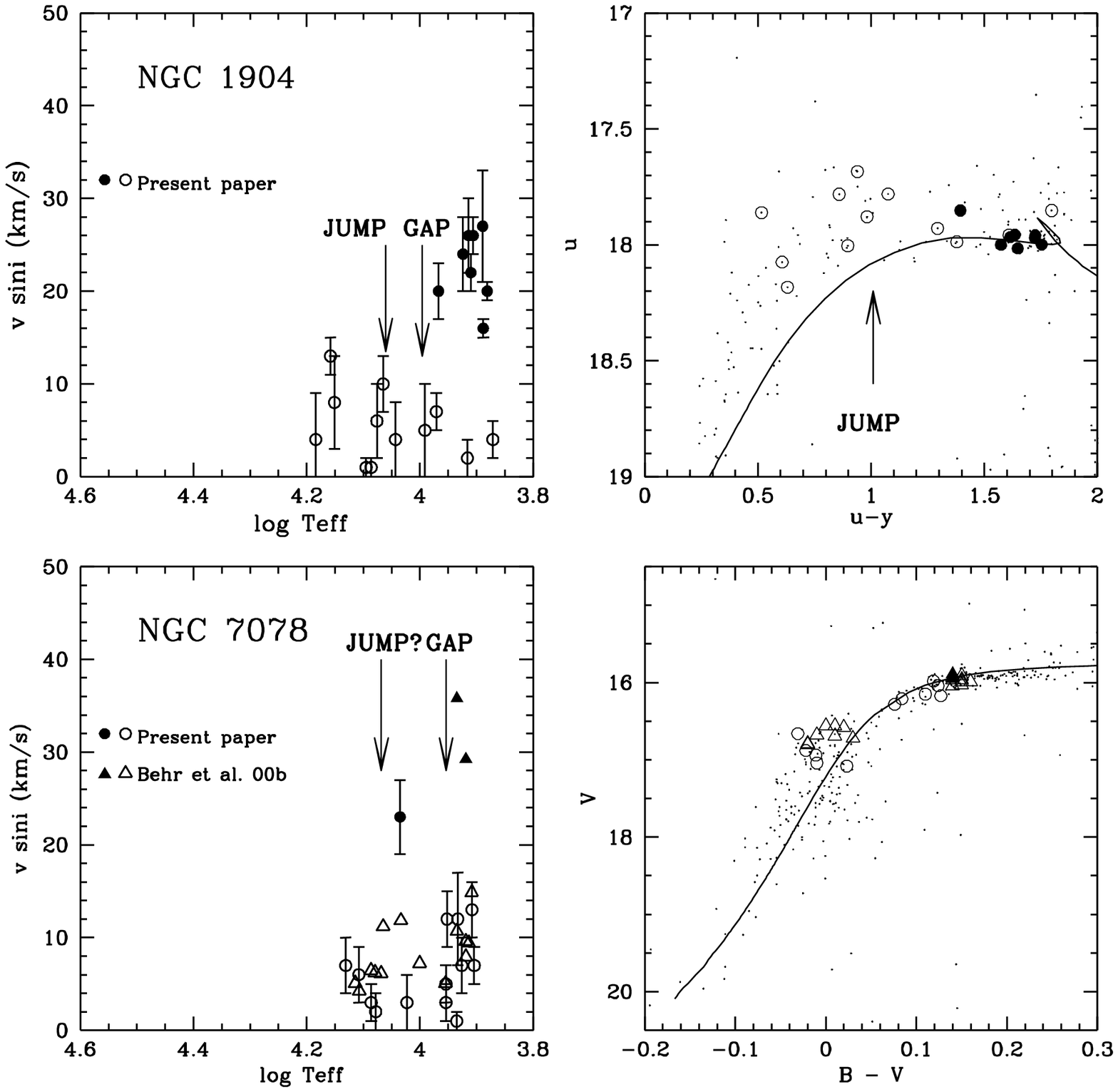}
\plotone{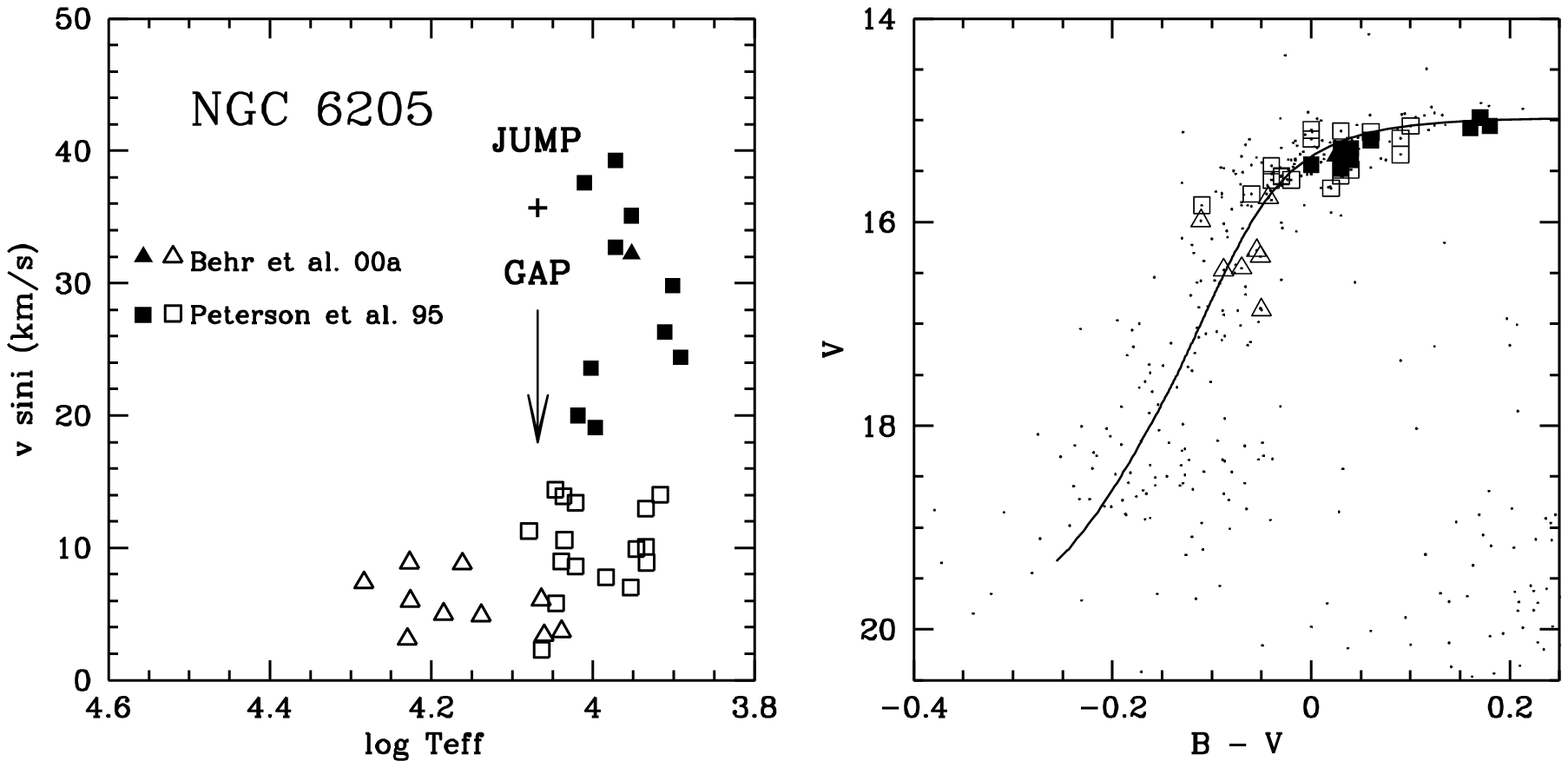}
\figcaption{As in Fig.\ 1 for M79, M15, and M13.
Full symbols show $v$sin$i\ge15$ km/s stars.}
\label{fast1}
\end{figure}


\begin{thebibliography}{}
\bibitem[Bedin(2000)]{bedin00} 
Bedin, L. R., Piotto, G., Zoccali, M., Stetson, P.B., Saviane, I.,
 Cassisi, S., \& Bono, G. 2000, \aap, 363, 159
\bibitem[Behr(1999)]{behr99}
Behr, B.B., Cohen, J. G., McCarthy, J.\ K., \& Djorgovski, S.\ G. 1999,
\apj, 517,L31
\bibitem[Behr(2000a)]{behr00a}
Behr, B.B.,Djorgovski, S. G., Cohen, J. G., McCarthy, J.\ K., C$\hat{o}$t$\acute{e}$, P., 
 Piotto, G., \& Zoccali, M. 2000, \apj, 528, 849 [B00a]
\bibitem[Behr(2000b)]{behr00b}
Behr, B.B., Cohen, J. G., \& McCarthy, J.\ K. 2000b, \apj, 531,L37 [B00b] 
\bibitem[Brown(2000)]{brown00}
Brown, T.M., Bowers, C.W., Kimble, R.A., \& Ferguson, H.C. 2000, \apj, 529, L89
\bibitem[Brown(2001)]{brown01}
Brown, T.M., Sweigart, A.V., Lanz, T., Landsman, W.B., Hubeny, I. 2001, \apj, 562, 368
\bibitem[Cassisi(1999)]{cas99}
Cassisi, S., Castellani, V., degl'Innocenti, S., Salaris, M., \& Weiss, A. 1999, 
 \aaps, 134, 103
\bibitem[D'Cruz(1996)]{dcruz96}
D'Cruz, N.\ L., Dorman, B., Rood, R. \ T. \& O'Connell, R. \ W. 1996,\apj, 466, 359
\bibitem[Dubath(1990)]{dub90} 
Dubath, P., Meylan, G., Mayor, M. et al.\ 1990, \aap, 239, 142
\bibitem[Ferraro(1998)]{fer98} 
Ferraro, F.~R., Paltrinieri, B., Fusi Pecci, F., Rood, R. T., \& Dorman, B. 1998,\apj, 500,
 311 [F98]
\bibitem[FusiPecci(1993a)]{fusi93a}
Fusi Pecci, F., Ferraro, F. R., Bellazzini, M., Djorgovski, S., Piotto, G., \& Buonanno, R.
 1993, \apj, 105, 1145
\bibitem[Gilliland(2000)]{gill00}
Gilliland, R., et al.\ 2000,\apj , 545, 47
\bibitem[Glaspey(1989)]{gla89}
Glaspey, J.W., Michaud, G., Moffat, A. F. J. \& Demers, S., 1989, \apj, 339, 926
\bibitem[Greggio(1990)]{greg90}
Greggio, L., \& Renzini, A. 1990, \apj, 364, 35
\bibitem[Greenstein(1967)]{gre67}
Greenstein, G.S., Truran, J.W., \& Cameron, A.G.W. 1967, Nature, 213, 871
\bibitem[Grundahl(1999)]{grun99}
Grundahl, F., Catelan, M., Landsman, W. B., Stetson, P. B., \& Andersen,
M.\  I. 1999,\apj , 524, 242 [G99]
\bibitem[Hill(1996)]{hill96}
Hill, R. S., Cheng, K. P., Smith, E. P., Hintzen, P. M. N.,
 Bohlin, R. C., \& O'Connell, R. W. 1996,\aj, 112,2909
\bibitem[Melo(2001)]{melo01}
Melo, C. H. F., Pasquini, L. \& De Medeiros, J.~R 2001, \aap, 375, 851
\bibitem[Mengel(1976)]{meng76}
Mengel, J.G., \& Gross, P.G. 1976, \apss, 41, 407
\bibitem[Moehler(2000)]{moeh00}
Moehler, S., Sweigart, A.V., Landsman, W. B. \& Heber, U., 2000, \aap, 360, 120
\bibitem[Peterson(1983)]{peter83}
Peterson, R. C. 1983,\apj,275,737
\bibitem[Peterson(1995)]{peter95}
Peterson, R. C., Rood, R.\ T., \& Crocker, D. A. 1995,\apj, 453; 214
\bibitem[Piotto(1999)]{pio99}
Piotto, G. et al.\ 1999,\aj, 118, 1737 [P99]
\bibitem[Piotto(2002)]{pio02}
Piotto, G. et al.\ 2002,\aap, submitted
\bibitem[Renzini(1977)]{renz77}
Renzini, A. 1977;in Advanced Stages in Stellar Evolution,Geneva Observatory, p.149 
\bibitem[Rich(1997)]{rich97}
Rich, R.\  et al.\ 1997,\apj , 484, L25
\bibitem[Rosenberg(2000)]{ros00}
Rosenberg, A., Piotto, G., Saviane, I., \& Aparicio, A. 2000,\aaps,144,5
\bibitem[Sills(2000)]{sill00}
Sills, A., \& Pinsonneault, M. H. 2000,\apj , 540, 489
\bibitem[Soker(1998)]{sok98}
Soker, N. 1998,\aj,116,1308
\bibitem[Sweigart(1997)]{sweig97}
Sweigart, A.V. 1997, in The 3$^{rd}$ Conference on Faint Blue Stars, eds.
  Philip, A.G.D., Liebert, J.W., \& Saffer, R.A, p. 3
\bibitem[Tonry(1979)]{ton79}
Tonry, J., \& Davis, M. 1979, \aj,84,1511
\bibitem[Vink(2000)]{vink00}
Vink, J.-S., de Koter, A., \& Lamers, H.J.G.L.M. 2000,\aap, 362, 295
\end{thebibliography}
\end{document}